\documentclass[12pt]{article}

\begin{document}

\author{James T. Wheeler}
\title{Quanta without quantization}
\date{Department of Physics, Utah State University, Logan, UT 84322}
\maketitle

\begin{abstract}
The dimensional properties of fields in classical general relativity lead to
a tangent tower structure which gives rise directly to quantum mechanical
and quantum field theory structures without quantization.

We derive all of the fundamental elements of quantum mechanics from the
tangent tower structure, including fundamental commutation relations, a
Hilbert space of pure and mixed states, measurable expectation values,
Schr\"{o}dinger time evolution, ``collapse'' of a state and the probability
interpretation.

The most central elements of string theory also follow, including an
operator valued mode expansion like that in string theory as well as the
Virasoro algebra with central charges.
\end{abstract}

\paragraph{Introduction}

The detailed structure of quantum systems follows by fully developing the
scaling structure of spacetime. Thus quantum systems - despite their Hilbert
space of states, operator-valued observables, interfering complex
quantities, and probabilities - are rendered in terms of classical spacetime
variables which simultaneously form a Lie algebra of operators on the space
of conformal weights.

This remarkable result follows from the \textit{tangent tower} structure
implicit in general relativity. We develop this structure, cite a central
theorem, then examine some properties of tensors over the tangent tower.
Finally, we apply these ideas to quantum mechanics and string theory,
showing how the core elements of both arise classically.

\paragraph{The weight tower and weight maps}

Consider a spacetime $(M,g)$ with dynamical fields, $\{\Phi _{A}\mid A\in
\mathcal{A}\}$ of various conformal weights and tensorial types. Under a
global change of units by $\lambda ,$ let
\begin{equation}
\Phi _{A}\rightarrow (\lambda )^{w_{A}}\Phi _{A}
\end{equation}
where $w_{A}$ is called the conformal weight of $\Phi _{A}$. Normally,
physicists simply insert these scale factors by hand when needed, without
mentioning the implicit mathematical structures their use requires, but
these structures turn out to be interesting and important.

The tower structure begins with the set of conformal weights, $W\equiv
\{w_{A}\mid A\in \mathcal{A}\},$ which must be closed under addition of any
two different elements. Possible sets include the reals $R,$ the rationals $%
Q,$ the integers $J,$ and the finite set $\{0,1,-1\}$. For most physical
problems we can choose the unit-weight objects to give $W=J$.

Next, define the equivalence relation
\begin{equation}
\Phi _{A}\cong \Phi _{B}\quad if\quad w_{A}=w_{B}\quad and\quad \exists \eta
\quad such\ that\quad \Phi _{A}=\eta \Phi _{B}
\end{equation}
which partitions the tangent space $TM$ into a tower of projective
Minkowski spaces, $PM$, one copy for each $n\in J=W.$ This partition
enlarges the linear transformation group of the tangent space into the
direct product of the Lorentz group and the group of \textit{weight maps}.
While the Lorentz transformations have their usual effect, general weight
maps act on conformal weight. To see that the group is a direct product,
consider the product of an $n$-weight scalar field with an $m$-weight vector
field$.$ Since the linear transformations preserve the Lorentz inner product
between arbitrarily weighted vectors, the resulting $(n+m)$-weight vector
field remains parallel to the original $m$-weight vector field under Lorentz
transformation. Therefore, Lorentz transformations map different weight
vectors in the same way.

We now investigate weight maps. Just as the only measurable Lorentz objects
are \textit{scalars}, the only measurable tangent tower quantities are
\textit{zero weight scalars}\footnote{%
Eg., we may measure the dimensionless ratio of the length of a table to the
length of a meter stick.}. Therefore, to readily form zero weight scalars,
we classify weight maps and tensors over $W$ by their conformal weights.

The generator of global scale changes, $D,$ determines the weights of fields
according to\footnote{%
In conformal geometries, $D$ generates dilations.}
\begin{equation}
D\Phi _{A}=w_{A}\Phi _{A}
\end{equation}
while the generators $M_{\alpha }$ of definite-weight maps satisfy
\begin{equation}
\lbrack D,M_{\alpha }]=n_{\alpha }M_{\alpha }
\end{equation}
where $n_{\alpha }\in J.$ Then $M_{\alpha }$ maps $n$-weight fields to $%
(n+n_{\alpha })$-weight fields, making the construction of $0$-weight
quantities straightforward.

The following theorem now holds [1]:

\paragraph{Theorem.}

Let $V$ be a maximally non-commuting Lie algebra consisting of exactly one
weight map of each conformal weight. Then $V$ is the Virasoro algebra with
central charge,
\begin{equation}
\lbrack M_{(m)},M_{(n)}]=(m-n)M_{(m+n)}+cm(m^{2}-1)\delta _{m+n}^{0}1
\end{equation}

The lengthy proof relies on explicit construction through a series of
inductive arguments. It is highly significant to note that \textit{the
real-projective tangent tower produces the same central charge for }$V$%
\textit{\ as unitarily-projective string theory, despite its classical
character}. This is the first concrete evidence that some phenomena widely
regarded as ``quantum'' can be understood from a classical standpoint.

\paragraph{Tensors over the weight tower}

Having understood the algebraic character of definite-weight weight maps, we
next look at the tensors they act on. Since Lorentz transformations decouple
from weight maps, the Lorentz and conformal ranks are independent. Thus
\begin{equation}
T_{n_{1}\cdots n_{s}}^{a_{1}\cdots a_{r}}:(PM)^{r}\otimes J^{s}\rightarrow R
\end{equation}
is a typical rank $(r,s)$ tensor. Weight maps act linearly on each label $%
n_{i}$.

Particularly relevant to quantum systems are $(0,1)$ tensors. With $\eta
_{k} $ a $k$-weight scalar, $D\eta _{k}=k\eta _{k},$ a general $(0,1)$
tensor is an indefinite-weight linear combination
\begin{equation}
\Phi =\sum_{k=-\infty }^{\infty }\phi _{k}\eta _{k}
\end{equation}
We immediately see the need for some convergence criterion. Imposing a norm
provides such a criterion, and with a norm these objects form a Hilbert
space, $H$. The simplest norm uses a continuous representation for the $%
(0,1) $ tensors defined by
\begin{equation}
\Phi (x)\equiv \sum_{n=-\infty }^{\infty }\phi _{n}e^{inx}.
\end{equation}
Clearly, $\Phi (x)$ exists only under appropriate convergence conditions,
provided by including as the $(0,1)$ tensors those vectors satisfying
\begin{equation}
\frac{1}{2\pi }\int_{-\pi }^{\pi }\Phi (x)\bar{\Phi}(x)<\infty
\end{equation}

Using the product
\begin{equation}
(MN)(x,y)=\int_{-\pi }^{\pi }M(x,z)N(z,y)dz
\end{equation}
we straightforwardly find the representations
\begin{equation}
D(x,y)=-i\frac{\partial }{\partial x}\delta (x-y)+c1
\end{equation}
for $D$ and
\begin{equation}
M_{(k)}(x,y)=e^{ikx}(-i\partial _{x}-k)\delta (x-y)
\end{equation}
for the $k$-weight Virasoro operator. In eq.(11), $D(x,y)$ requires the
central distribution $c$ to cancel surface terms from the product integral.
Such terms are an artifact of the continuous representation.

\paragraph{Physical effects of the tangent tower}

Now we describe physical effects of the tangent tower. We replace the usual $%
(r,0)$ tensors of quantum mechanics and quantum field theory by $(r,1)$ or $%
(r,2)$ tensors. In the remaining two sections we do this in detail for
quantum mechanics, then briefly for string theory.

\paragraph{Conformal weight and quantum mechanics}

The tangent tower underpinning of all axiomatic features of quantum
mechanics now follows immediately: commutation relations of operator-valued
position and momentum vectors, a Hilbert space of pure and mixed states,
measurable expectation values, Schr\"{o}dinger time evolution, ``collapse''
of a state and the probability interpretation.

First, consider canonical commutators. Since $w_{x}=1$ and $w_{p/h}=-1$,
canonical coordinates lie in the subalgebra determined by $W_{0}=$ $%
\{0,1,-1\}\subset W.$ We temporarily consider operators in this subspace,
replacing symplectic coordinates $Q^{A}=(q^{i},\pi _{j})$ by
weight-map-valued 6-vectors $\hat{Q}^{A}=(\hat{q}^{i},\hat{\pi}_{j})\in
T_{nm}^{a}$ forming the Lie algebra
\begin{equation}
\lbrack \hat{Q}^{A},\hat{Q}^{B}]=c^{AB}1
\end{equation}
where projective representation allows arbitrary central charges $c^{AB}$.
The only invariant antisymmetric symplectic tensor is the symplectic 2-form
\begin{equation}
\Omega ^{AB}=\left(
\begin{array}{ll}
0 & -\delta _{j}^{i} \\
\delta _{i}^{j} & 0
\end{array}
\right)
\end{equation}
so we set
\begin{equation}
\lbrack \hat{Q}^{A},\hat{Q}^{B}]=\Omega ^{AB}1
\end{equation}
Furthermore, the natural symplectic metric
\begin{equation}
K^{AB}=\left(
\begin{array}{ll}
0 & \delta _{j}^{i} \\
\delta _{i}^{j} & 0
\end{array}
\right)
\end{equation}
may be diagonalized by a symplectic transformation to new variables $\hat{R}%
^{A}=(\hat{X}^{i},\hat{P}_{j})$ such that
\begin{equation}
\tilde{K}^{AB}=\left(
\begin{array}{ll}
\delta _{ij} &  \\
& -\delta ^{ij}
\end{array}
\right)
\end{equation}
or equivalently $\hat{R}^{\prime A}=(\hat{X}^{i},i\hat{P}_{j})$ with
\begin{equation}
\tilde{K}^{\prime AB}=\left(
\begin{array}{ll}
\delta _{ij} &  \\
& \delta ^{ij}
\end{array}
\right)
\end{equation}
The zero signature of the symplectic metric \textit{requires} an imaginary
unit in relating $\hat{R}^{\prime A}$ to $\hat{P}_{j}$ because the physical $%
\hat{X}^{i}$ and $\hat{P}_{j}$ have the \textit{same} metric, $+\delta
_{ij.} $ The projective algebra of $\hat{X}^{i}$ and $\hat{P}_{j}$ is the
canonical one

\begin{eqnarray}
\lbrack \hat{X}^{i},\hat{P}^{j}] &=&i\delta ^{ij}1 \\
\lbrack \hat{X}^{i},\hat{X}^{j}] &=&[\hat{P}^{i},\hat{P}^{j}]=0
\end{eqnarray}
Thus, ``quantum'' $(\hat{X}^{i},\hat{P}^{j})$ commutators follow from
classical scaling considerations by replacing classical variables with
weight tower operators in the $\{0,1,-1\}$ subalgebra, and using the natural
symplectic structure.

\smallskip

Other quantum structures follow easily. Thus, definite weight scalars
replace pure quantum states, while indefinite weight objects such as $\Phi
(x)\in H$ replace mixed states. The construction of expectation values is
simply a rule to generate a $0$-weight scalar. The rule works by matching
elements of $H$ with their complex conjugates but the definition of $\Phi
(x) $ above translates this into the manifestly $0$-weight sum
\begin{eqnarray}
\langle \Phi ,\Psi \rangle &=&\frac{1}{2\pi }\int_{-\pi }^{\pi }\Phi (x)\bar{%
\Psi}(x)dx \\
&=&\sum \phi _{n}\psi _{-n}
\end{eqnarray}

Finally, consider the time evolution of $\Phi (x).$ In standard quantum
mechanics, states evolve \textit{continuously} via the Schr\"{o}dinger
equation and \textit{discontinuously} during measurement. Conformal
properties account for both processes.

\textit{Continuous evolution} of weighted quantities is described by
parallel transport, where the metric-compatible covariant derivative $\nabla
$ must be augmented by the $(1,2)$ gauge vector $W_{\mu }$ of the \textit{\
full }tangent tower symmetry. With the weight-map-valued 1-form $\mathbf{W}%
=W_{\mu }\mathbf{d}x^{\mu }$ exact, the geometry remains Riemannian, but
having $\mathbf{W,}$ we can write expressions which are manifestly
scale-invariant, even locally. Parallel transport of a $(0,1)$ tensor obeys
\begin{equation}
u^{\mu }\nabla _{\mu }\Phi (x^{\nu };x)=-u^{\mu }W_{\mu }\Phi (x^{\nu };x)
\label{Schrodinger}
\end{equation}
where $u^{\mu }\nabla _{\mu }$ reduces to $\frac{d}{d\tau }$ in flat
spacetime while $u^{\mu }W_{\mu }\Phi $ gives the action of a weight map $%
\mathcal{H}\equiv u^{\mu }W_{\mu }$ on the weight superposition $\Phi
(x^{\nu };x).$ Identifying $\mathcal{H}$ as the Hamiltonian operator [2],
eq.(22) becomes the Schr\"{o}dinger equation. The free-particle form for $%
\mathcal{H}$ is the zero-weight quantity
\begin{equation}
\mathcal{H}=\left( \frac{d\hat{X}^{j}}{d\tau }\right) \left( i\hat{P}%
_{j}\right)
\end{equation}
so the imaginary unit enters correctly.

\textit{Discrete evolution} and the probability interpretation are natural
outcomes of measuring \textit{properties} with definite conformal weight.
Such measurement involves projecting a superposition $\Phi (x)$ into an
object with, eg., units of length (giving $\phi _{1}$), inverse area (giving
$\phi _{-2}$), etc. What we actually measure are dimensionless ratios, $%
\frac{\phi _{1}}{L},$ $L^{2}\phi _{-2},$ etc., with $L$ some standard unit
of length.

Therefore \textit{the rules of quantum mechanics have natural classical
interpretations in terms of the scale-invariance properties of spacetime.}

\paragraph{Conformal weight and quantum field theory}

Just as field theory emerges as the limiting case of multiparticle dynamics,
and quantum field theory emerges as a blend of quantum mechanics and special
relativity [3], we can imagine retracing the preceding arguments to derive
the principles of quantum field theory. After all, the quantum state \textit{%
is} a field, even in quantum mechanics. Thus, we may anticipate a classical
tangent tower interpretation of quantum field theory. As striking as this
conjecture seems, we now demonstrate a more striking claim: the tangent
tower contains the essential elements of string theory [4].

Minor manipulations of a definite weight $(1,2)$ tensor $\alpha _{(k)}^{\mu
}(x,y)$ found by writing the $\delta $-function in $D$ as
\begin{equation}
\delta (x-y)=\frac{1}{2\pi }\sum_{n=-\infty }^{\infty }e^{in(x-y)}
\end{equation}
lead to
\begin{equation}
\alpha _{(k)}^{\mu }(x,y)=(2\pi )^{-1}e^{ikx}\sum\limits_{m}\left[ \alpha
_{(k)m}^{\mu }e^{im(x-y)}+\tilde{\alpha}_{(k)m}^{\mu }e^{im(x+y)}\right]
\end{equation}
with the $0$-weight $(1,2)$ tensors given by the Fourier series
\begin{equation}
\alpha ^{\mu }(x,y)=(2\pi )^{-1}\sum\limits_{m}\left[ \alpha _{m}^{\mu
}e^{im(x-y)}+\tilde{\alpha}_{m}^{\mu }e^{im(x+y)}\right]
\end{equation}
This expression has the form of a string mode expansion, with each mode
being a Hilbert space operator. Viewing the two modes as left and right
moving waveforms, gives the abstract $(x,y)$ space a Lorentz norm. Under
mild assumptions [5], time-oriented embeddings of the abstract space into
spacetime exist, and have the properties of string world sheets.

Nothing is more central to the theory of quantized string than these mode
operators. Through the Virasoro algebra they determine the physical states.
Mode operators give a representation of the Poincar\'{e} generators thereby
classifying those states by mass and spin. Ultimately this gives a gauge
theory of a massless spin-2 mode governed at lowest order by the
Einstein-Hilbert action. Additionally, the Virasoro central charge
determines the critical dimension of spacetime. The existence of heterotic
and supersymmetric representations of the symmetry eliminates tachyons,
guarantees fermions and provides a large internal nonabelian gauge symmetry.

All of these string properties may now be studied as properties of $(1,2)$
tensors on the conformal tangent tower of spacetime.\pagebreak

\paragraph{References}

\begin{enumerate}
\item[{\lbrack 1]}]  Wheeler, J. T., String without strings, in preparation.

\item[{\lbrack 2]}]  Wheeler, J. T., Phys. Rev. D\textbf{44} (1990);
Wheeler, J. T, Proceedings of the Seventh Marcel Grossman Meeting on General
Relativity , R. T. Jantzen and G. M. Keiser, editors, World Scientific,
London (1996) pp 457-459.

\item[{\lbrack 3]}]  Weinberg, S., \textit{The quantum theory of fields,}
Cambridge University Press (1995).

\item[{\lbrack 4]}]  Green, M. B., J. H. Schwarz and E. Witten, \textit{\
Superstring theory}, Cambridge University Press (1987).

\item[{\lbrack 5]}]  Kuchar, K. V. and C. G. Torre, J. Math. Phys. \textbf{30%
} (8), August 1989.
\end{enumerate}

\end{document}